\documentclass[conference]{IEEEtran}
\ifCLASSINFOpdf
\else
\fi
\usepackage{url}
\usepackage{amsmath}
\usepackage{algpseudocode}
\usepackage{algorithm}
\usepackage{hhline}
\usepackage{graphicx}
\graphicspath{ {figures/} }
\usepackage{svg}

\begin{document}
%
\title{\textbf{MOSQUITO:} Covert Ultrasonic Transmissions between Two Air-Gapped Computers using  
	\textit{Speaker-to-Speaker} Communication}

\author{\IEEEauthorblockN{Mordechai Guri, Yosef Solewicz, Andrey Daidakulov, Yuval Elovici}
\IEEEauthorblockA{Ben-Gurion University of the Negev, Israel\\Cyber-Security Research Center\\
gurim@post.bgu.ac.il; yosef.solewicz@gmail.com; daydakul@post.bgu.ac.il; elovici@post.bgu.ac.il}}


%


\maketitle

\begin{abstract}
In this paper we show how two (or more) air-gapped computers in the same room, equipped with passive speakers, headphones, or earphones can covertly exchange data via ultrasonic waves. Microphones are not required. Our method is based on the capability of a malware to exploit a specific audio chip feature in order to reverse the connected speakers from output devices into input devices - unobtrusively rendering them microphones \cite{guri17speake}. We discuss the attack model and provide technical background and implementation details. We show that although the reversed speakers/headphones/earphones were not originally designed to perform as microphones, they still respond well to the near-ultrasonic range (18kHz to 24kHz). We evaluate the communication channel with different equipment, and at various distances and transmission speeds, and also discuss some practical considerations. Our results show that the speaker-to-speaker communication can be used to covertly transmit data between two air-gapped computers positioned a maximum of nine meters away from one another. Moreover, we show that two (microphone-less) headphones can exchange data from a distance of three meters apart. This enables 'headphones-to-headphones' covert communication, which is discussed for the first time in this paper.
\end{abstract}


%
\IEEEpeerreviewmaketitle

\section{Introduction}
Two (or more) computers in the same room are considered to be separated by an 'air-gap' if there is no physical or logical connection between them. In the context of cyber security, this measure is taken in order to ensure strict isolation between nearby computers. A common scenario involves two computers in the same room, where each computer is connected to a separate network of the organization. The air-gap separation ensures that data cannot be exchanged between the two networks, and more specifically, in a situation in which two computers have been compromised with a malware, data cannot be sent from one computer to the other and vice versa.

\subsection{Speaker-to-Microphone Covert Channel}
Despite the high degree of isolation provided by air-gapping, it doesn't provide a hermetic solution. It is known that the air-gap between two computers in the same room can be 'bridged' if the two computers are equipped with speakers and microphone \cite{hanspach2014covert,carrara2014acoustic}. That is, the two computers can covertly exchange data via inaudible sound waves. In this type of communication, one computer transmits the data to the other via high frequency sound (usually at 18kHz or higher), using its loudspeaker.
The receiver computer uses its microphone to receive the data. The speaker-to-microphone communication described above is mainly relevant for laptops, which have built-in speakers and microphones. Hence, previous research on this covert channel has primarily focused on laptops \cite{hanspach2014covert}. 
\subsection{Microphone-less Environments}
The speaker-to-microphone covert channel has one main drawback: in many real-life IT environments, microphones are not available to the attacker. The common cases include:
\begin{itemize}
	\item \textbf{Desktop workstations}. Unlike laptops which have integrated microphones, desktop workstations are not always connected with an external microphone. 
	\item \textbf{Secure environments}. In secure environments, microphones in desktop computers may be prohibited (or disconnected) to avoid the risk of eavesdropping. In secure environments, microphones may be forbidden in order to maintain an 'audio-gap' between computers. Elimination of microphones is an effective defense against the speaker-to-microphone covert channel discussed above \cite{Jumpingt83:online}.
	\item \textbf{Disabled/muted microphones}. A computer (desktop workstation or laptop) may be equipped with a microphone, which at some point was disabled, muted (with a physical 'off' button), or taped  \cite{WhyhasMa55:online}. This typically occurs when the user wants to increase security and ensure confidentiality.       
\end{itemize}

Consequently, the speaker-to-microphone covert channel limits the attacker's abilities, allowing the attacker to operate only in environments where microphones  \textit{are} present and enabled.  

\subsection{Speaker-to-Speaker Covert Channel}
In this paper we show how the air-gap between two isolated computers can be bridged in 'speakers-only' environments. That is, where two computers in the same room are not equipped with microphones but are equipped with different types of output devices: (microphone-less) headphones, (microphone-less) earphones/earbuds, or passive speakers. Our method is based on the capability of a malware to transform a computer speaker from an output device into an input device - inconspicuously changing its role from speaker to microphone and vice versa  \cite{guri17speake}. The two computers can then be used to send data (by using the speakers) and receive data (by using the transformed speakers) via inaudible sound. \\ \\
The contribution of this paper is as follows: 
\begin{itemize}
	\item \textbf{Attack model.} We extensively discuss the speaker-to-speaker communication attack model of bridging the air-gap between two desktop computers. We also discuss and evaluate different types of speakers and headphones and their response to the ultrasonic range.
	\item \textbf{Speakers-to-headphones.}  We discuss and evaluate the never discussed before threat of the speaker-to-headphones and headphones-to-headphones communication channel. We show that two pairs of headphones can establish covert ultrasonic communication from a distance of three meters apart.
	\item \textbf{Evaluation.} We evaluate the speaker-to-speaker ultrasonic covert channel. In particular, we evaluate the acoustic response of passive speakers, headphones, and earphones to the near-ultrasonic range, when transformed into microphones (recall that such speakers are not designed to function as input devices).
	\item \textbf{Transmission protocol.} We provide a protocol stack designed for speaker-to-speaker communication. In this covert channel, the two computers must synchronize and change the speakers' roles (from speakers to microphones and vice versa) during the communication. We developed an appropriate protocol to handle this mutual communication.
	\item \textbf{Practical considerations.} We discuss and evaluate practical considerations regarding this covert channel, particularly, the effect of environmental noise on the channel's quality. We also discuss the position of the speakers and its effect on the signal strength.\\  
\end{itemize}
The rest of this paper is organized as follows: Technical background is provided in Section \ref{sec:tech}. Related work is presented in Section \ref{sec:related}. The attack is discussed in Section \ref{sec:attack}.   Communication details are provided in Section \ref{sec:communication}. Section \ref{sec:analysis} describes the analysis and evaluation results. Countermeasures are discussed in Section \ref{sec:counter}. We conclude in Section \ref{sec:conclusion}.

\section{Technical Background}
\label{sec:tech}
In this section, we provide the technical background necessary to understand the attack itself. An essential part of the speaker-to-speaker covert channel is malware's ability to record audio signals through the speakers/headphones/earphones connected to the computer. In the following subsection, we describe this issue and discuss its limitations.

\subsection{Speaker Reversibility}
A speaker aims at amplifying audio streams out, but it can actually be viewed as a microphone working in reverse mode: a loudspeaker converts electric signals into a sound waveform, while a microphone transforms sounds into electric signals. More technically, speakers use the changing magnetic field induced by electric signals to move a diaphragm in order to produce sounds. Similarly, in microphone devices, a small diaphragm moves through a magnetic field according to a sound's air pressure, inducing a corresponding electric signal  \cite{ballou2013handbook}. This bidirectional mechanism facilitates the use of a simple speaker as a feasible microphone simply by plugging it into a microphone jack. It should be clear that in practice, speakers were not designed to perform as microphones, and the recorded signals will be of low quality. 

\subsection{Jack Retasking}
Interestingly, the audio chipsets in modern motherboards and sound cards include an option to change the function of an audio port at the software level, a type of audio port programming sometimes referred to as 'jack retasking'. This option is available on most audio chipsets (e.g., Realtek's audio chipsets) integrated into PC motherboards today. Jack retasking, although documented in the technical specifications, is not well-known \cite{Turnyour67:online}. For an in-depth technical discussion on malicious retasking of an audio jack, from the hardware to the operating system level, we refer the interested reader to the following previous work \cite{guri17speake}.\\

The fact that loudspeakers, headphones, earphones, and earbuds are physically built like microphones, coupled with the fact that an audio port's role in the PC can be altered programmatically, changing it from output to input, creates a vulnerability which can be abused by attackers. A malware can stealthily reconfigure the headphone jack from a line out jack to a microphone jack. As a result, the connected output device can function as a pair of recording microphones, thereby rendering the computer a recording device - even when the computer does not have a connected microphone.

\subsection{Passive speakers, Headphones and Earphones}
The reversibility of speakers poses a limitation, in that the speaker must be passive (unpowered), without amplifier transitions. In the case of an active (externally powered) speaker, there is an amplifier between the jack and the speaker; hence, the signal will not be passed from the output to the input side \cite{duncan1996high}. Headphones, earphones, and earbuds are built from a pair of passive speakers, and hence, are always reversible. However, most PC loudspeakers today have an internal amplifier \cite{Powereds45:online}. Passive speakers mainly exist in legacy and intercom systems \cite{NSTISSAM75:online}. 

\begin{table}[t]
	\centering
	\caption{Audio output devices and their reversibility}
	\label{table1-list}
	\begin{tabular}{ll}
		\hline
		Device            & Reversible \\ \hhline{==}
		Active speaker    & No         \\
		Passive speaker   & Yes        \\
		Headphones        & Yes        \\
		Earphones/earbuds & Yes        \\ \hline
	\end{tabular}
\end{table}

Table \ref{table1-list}. lists the audio output devices and their reversibility. Active speakers are not reversible, and hence, can only act as the transmitting side in our covert channel. The receiving side must be a computer connected with passive speakers, headphones, or earphones.

\section{Related Work}
\label{sec:related}
Air-gap covert channels are special covert channels, which enable communication from air-gapped computers - mainly for a purpose of data exfiltration. They can be classified into five main categories: electromagnetic, magnetic, acoustic, thermal, and optical. 

\subsection{Electromangetic}
In the past twenty years, several studies have proposed the use of electromagnetic emanation from computers for covert communication. Kuhn  showed that it is possible to control the electromagnetic emissions from  computer displays \cite{kuhn1998soft}. Using this method, a malicious code can generated radio signals and modulate data on top of them. In 2014, Guri et al demonstrated AirHopper \cite{guri2014airhopper,guri2017bridging}, malware that exfiltrate data from air-gapped computers to a nearby smartphone via FM signals emitted from the screen cable. Later on Guri et al also demonstrated GSMem \cite{guri2015gsmem}, malware that leaks data from air-gapped computers to nearby mobile-phones using cellular frequencies generated from the buses which connect the RAM and the CPU. In 2016, Guri et al showed USBee, a malware that uses the USB data buses to generate electromagnetic signals from a desktop computer \cite{guri2016usbee}. 
\subsection{Magnetic}
In 2018, Guri et al presented ODINI \cite{guri2018odini}, a malware that can exfiltrate data from air-gapped computers via low frequency magnetic signals generated by the computer's CPU cores. The magnetic fields bypass Faraday cages and metal shields. Guri et al also demonstrated MAGNETO \cite{guri2018magneto}, which is a malware that leak data from air-gapped computers to nearby smartphones via magnetic signals. They used the magnetic sensor integrated in smartphones to receive covert signals. Matyunin suggested using magnetic head of hard disk drives to generate magnetic emission, which can be received by a nearby smartphone magnetic sensor \cite{matyunin2016covert}.

\subsection{Optical}
Several studies have proposed the use of optical emanation from computers for covert communication. Loughry introduced the use of PC keyboard LEDs to encode binary data \cite{loughry2002information}. In 2017, Guri et al presented LED-it-GO, a covert channel that uses the hard drive indicator LED in order to exfiltrate data from air-gapped computers \cite{Guri2017}. Guri et al also presented a method for data exfiltration from air-gapped networks via router and switch LEDs \cite{guri2017xled}. Data can also be leaked optically through fast blinking images or low contrast bitmaps projected on the LCD screen \cite{guri2016optical}.  In 2017, Guri et al presented aIR-Jumper, a malware that uses the security cameras and their IR LEDs to communicate with air-gapped networks remotely \cite{guri2017air}. 

\subsection{Thermal}
In 2015, Guri et al introduced  BitWhisper \cite{guri2015bitwhisper}, a thermal covert channel allowing an attacker to establish bidirectional communication between two adjacent air-gapped computers via temperature changes. The heat is generated by the CPU/GPU of a standard computer and received by temperature sensors that are integrated into the motherboard of the nearby computer.

\subsection{Acoustic}
In acoustic covert channels, data is transmitted via inaudible, ultrasonic sound waves. Audio based communication between computers was reviewed by Madhavapeddy et al. in 2005 \cite{madhavapeddy2005audio}. In 2013, Hanspach \cite{hanspach2014covert} used inaudible sound to establish a covert channel between air-gapped laptops equipped with speakers and microphones. Their botnet established communication between two computers alocated ~19 meters apart and can achieve a bit rate of 20 bit/sec. Deshotels \cite{deshotels2014inaudible} demonstrated the acoustic covert channel with smartphones, and showed that data can be transferred up to 30 meters away. In 2013, security researchers claimed to find a malware (dubbed BadBios) which communicates between two instances of air-gapped laptops via the integrated speakers and microphones using ultrasonic signals \cite{Meetbad27:online}. 
\\

\textbf{Speaker-less computers.} All of the acoustic methods presented above require speakers.  In 2016, Guri et al introduced Fansmitter, a malware which facilitates the exfiltration of data from an air-gapped computer via noise intentionally emitted from the PC fans \cite{guri2016fansmitter}. In this method, the transmitting computer does not need to be equipped with audio hardware or an internal or external speaker. Guri et al also presented DiskFiltration a method that uses the acoustic signals emitted from the hard disk drive (HDD) moving arm to exfiltrate data from air-gapped computers \cite{guri2017acoustic}.
\\

\textbf{Microphone-less computers.} The attack presented in the current paper is relevant to environments in which the computers are not equipped with microphones, a common setup seen in many IT environments. Guri et al presented Speake(a)r \cite{guri17speake} a malware that covertly turns the headphones, earphones, or simple earbuds connected to a PC into a pair of eavesdropping microphones when a standard microphone is not present, muted, taped, or turned off. They discuss technical details of this type of attack from the hardware to operating system level. However, the work of Guri et al in \cite{guri17speake} focuses on the threat of conversation eavesdropping and did not discuss the threat of ultrasonic covert channel. Lee et al. evaluate the various acoustic (non-covert) channels, and suggested  establishment of communication between two loudspeakers. However, with inaudible range (above 18kHz) they achieved a limited distance of 10 centimeters \cite{lee2015various}. They use passive loudspeakers and didn't evaluate headphones, earphones or earbuds for the transmission and reception. As we noted, most loudspeakers connected to PCs today have an integral amplifier which prevents passing any signal from output to input.
\\
\begin{figure*}[]
	\centering
	\includegraphics[width=0.7\textwidth,trim=2cm 3.4cm 2cm 3cm]{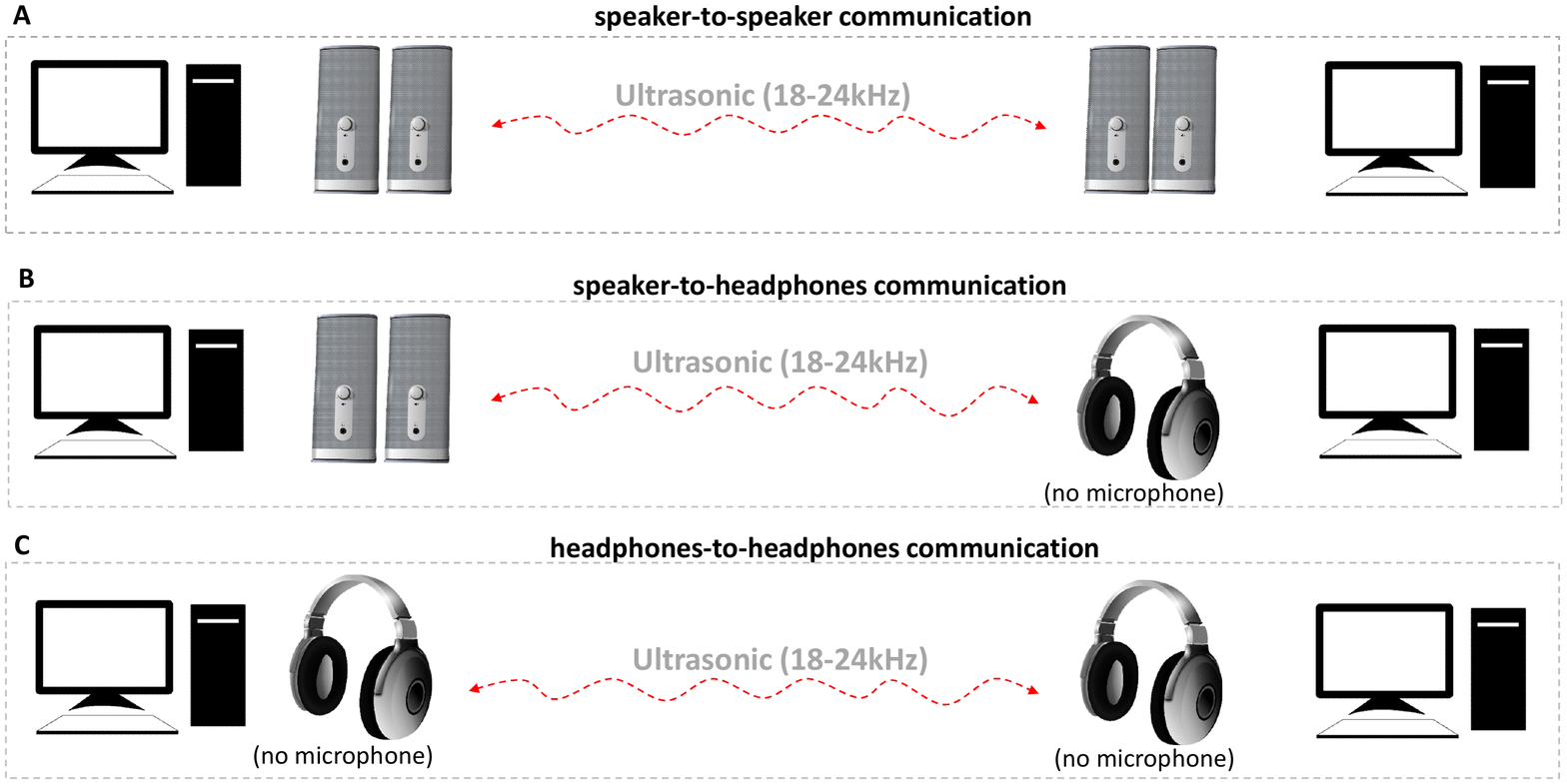}
	\caption{The three communication scenarios of the proposed covert channel. (A) speaker-to-speaker communication, (B) speaker-to-headphones communication, and (C) headphones-to-headphones communication}
	\label{fig:three}
\end{figure*}

\begin{table}[]
	\centering
	\caption{Summary of existing air-gap covert channels}
	\label{table-related}
	\begin{tabular}{ll}
		\hline
		Type               & Method                                                                                                                                                                                          \\ \hhline{==}
		Electromagnetic    & \begin{tabular}[c]{@{}l@{}}AirHopper \cite{guri2014airhopper,guri2017bridging} (FM radio) \\ GSMem \cite{guri2015gsmem} (cellular frequencies)  \\ USBee \cite{guri2016usbee} (USB bus emission) \\ Funthenna \cite{funtenna86:online} (GPIO emission) \end{tabular}                                                                   \\ \hline
		Magnetic           & \begin{tabular}[c]{@{}l@{}}MAGNETO \cite{guri2018magneto} (CPU-generated \\ magnetic fields)\\ ODINI \cite{guri2018odini} (Faraday shields bypass) \\ Hard-disk-drive \cite{matyunin2016covert} \end{tabular}                                                                                            \\ \hline
		Acoustic           & \begin{tabular}[c]{@{}l@{}}Fansmitter \cite{guri2016fansmitter} (computer fan noise) \\ DiskFiltration \cite{guri2017acoustic} (hard disk noise) \\ Ultrasonic \cite{hanspach2014covert,carrara2014acoustic}\\ MOSQUITO (speaker-to-speaker)\end{tabular} \\ \hline \\
		Thermal            & BitWhisper  \cite{guri2015bitwhisper} (heat emission)                                                                                                                                                                           \\\\ \hline 
		Optical            & \begin{tabular}[c]{@{}l@{}}LED-it-GO \cite{Guri2017} (hard drive LED) \\ VisiSploit \cite{guri2016optical} (invisible pixels) \\ Keyboard LEDs \cite{loughry2002information}\\ Router LEDs \cite{guri2017xled}\end{tabular}                            \\ \hline 
		Optical (infrared) & aIR-Jumper \cite{guri2017air} (security cameras \& infrared)                                                        \\ \hline
	\end{tabular}
\end{table}
Table \ref{table-related}. summarizes the existing air-gap covert channels.

\section{Attack}
\label{sec:attack}
In the attack scenario, two or more computers are located in the same room - separated by an air-gap. That is, there is no physical or logical network connection between the two computers. The computers are not equipped with microphones but are equipped with output devices: active speakers, passive speakers, headphones, or earbuds. Fig. \ref{fig:three} illustrates three scenarios of the proposed covert channel. (A) speaker-to-speaker communication, (B) speaker-to-headphones communication, and (C) headphones-to headphones communication.

We distinguish between two types of communication.
\begin{itemize}
	\item \textbf{Two computers, one-way communication.} In this case, two air-gapped computers in the same room establish unidirectional communication. This is the simplest case, where one computer is a transmitter and the other is a receiver. In this case, the transmitter is not necessarily equipped with a reversible speaker (e.g., it might be connected to an active loudspeaker).
	\item \textbf{Two computers, bidirectional communication.} In this case, two air-gapped computers in the same room establish bidirectional communication. In this case, each of the computers is a transmitter and a receiver. In this case, both computers are equipped with reversible speakers.
\end{itemize}

\subsection{Malware}
The communicating computers are infected with a malware. The malware has three operational components, described below.
\begin{itemize}
	
	\item \textbf{Jack retasking.} Reversing the output audio jacks into input jacks, effectively turning the playing devices to microphones. This technique is described in detail in \cite{guri17speake}. 
	\item \textbf{Synchronization.} Synchronizing between the sender and the receiver. This component is essential for a bidirectional communication.  By using the synchronization, the malware determines when the speaker should be used as a speaker and when it should be reversed to a microphone. 
	\item \textbf{Transmission and reception.} Transmitting and receiving the data. This component performs the modulation of the data over ultrasonic waves and its demodulation back to binary data. It also includes the bit framing and the transmission protocol.
\end{itemize}
\subsection{Air-Gap Communication}
The attack presented in this paper allow attackers to transmit data between two computers. For example, when one computer is Internet connected and the other is an isolated, air-gapped computer. In the initial phase, the two computers must be infected with a malware. Note that  it has been shown that attackers can successfully compromise air-gapped networks by using complex attack vectors, such as supply chain attacks, malicious insiders, and social engineering \cite{maybury2005analysis,TrumpPut87:online,abraham2010overview}. For example, in 2017 WikiLeaks published a reference to a hacking tool dubbed 'Brutal Kangaroo,' used to infiltrate air-gapped computers via USB drives \cite{Wikileak92:online}. When an employee of the organization inserted an infected USB drive into the air-gapped computer, a malicious code was executed.

Having established a foothold in both computers,
the attacker may bridge the air-gap between the internal and external networks using the speaker-to-speaker covert channel. The attacker can then exfiltrate information to the Internet connected computer (e.g., documents, passwords and encryption keys). Alternatively, the attacker may communicate with the isolated network by issuing commands and receiving responses.

\section{Communication}
\label{sec:communication}
In this section, we present the design and implementation of the speaker-to-speaker communication. We discuss the detection and synchronization protocol and present the data modulation and encoding scheme. For this discussion, we assume that there are two computers (or 'nodes') denoted as \textbf{A} and \textbf{B}. We present a generic protocol which assumes that both \textbf{A} and \textbf{B} are connected with a passive speaker or headphones/earphones. At the end of this section, we discuss a case in which only one of the computers is equipped with a reversible speaker, allowing only unidirectional (rather than bidirectional) communication. Note that for simplicity we present the basic case with only two communicating peers.

\subsection{Protocol Stack}
The approach taken by other research on the ultrasonic covert channel is to use the existing implementation of protocol stacks originally designed for acoustic (non-ultrasonic) communication (e.g., \cite{hanspach2014covert,mccoy2010janus}). In this paper, we choose to implement our own light, stripped-down audio protocol stack for the evaluation of the covert channel.

\subsection{The Near-Ultrasonic Range}
Human hearing is limited to sound waves of ~20kHz. In covert channel research it is acceptable to classify the range above 18kHz as practically inaudible for adults \cite{deshotels2014inaudible,vitello2006ring}. In 2016, a group of researchers performed in-depth analysis on the emerging threat of ultrasonic cross-device tracking (uXDT). They found that ultrasound beacons
(uBeacons) at a range of 18kHz to 20kHz are embedded into websites and TV advertisements \cite{mavroudis2017privacy}. The beacons are then picked up apps installed on nearby smartphones. Accordingly, in this paper, we consider the frequency range of 18kHz to 24kHz acceptable for the covert communication. 

\subsection{Data Modulation}
In audio frequency-shift keying (AFSK) digital data is represented by changes in the frequency of an audio tone. AFSK is used to transmit binary data over radio and telephony systems. For the data transmission we implemented binary frequency-shift keying (B-FSK) modulation.  In B-FSK the data is represented by a change in the frequency of a carrier wave. In our case, two different audio frequencies $f_{0}$ and $f_{1}$ in the range of 18kHz to 24kHz represent two different symbols '0' and '1.'

\begin{figure*}[t]
	\centering
	\includegraphics[width=\textwidth, ,trim=0cm 0.5cm 0cm 0cm]{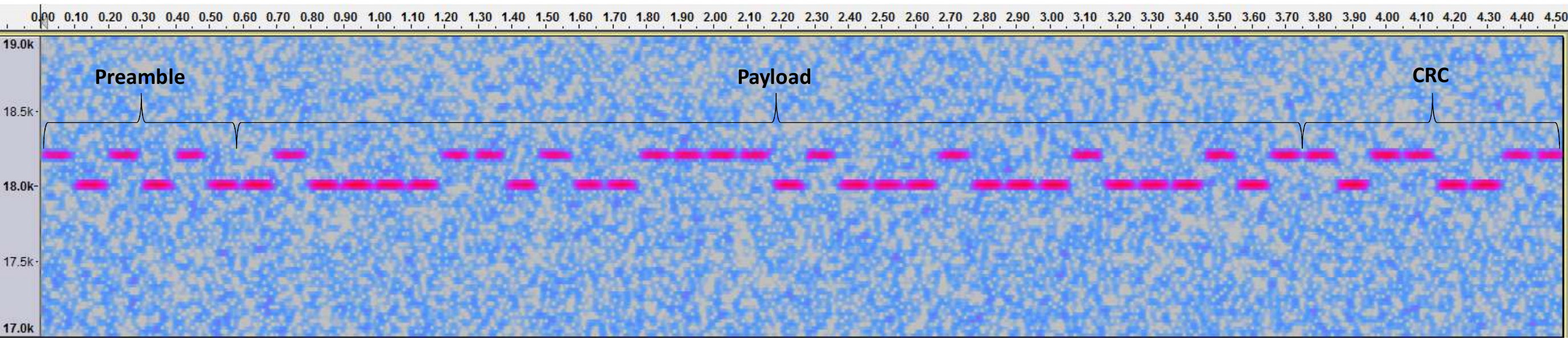}
	\caption{The spectrogram of the 46 bit frame (preamble, payload, and CRC) transmitted at 10 bit/sec using B-FSK modualtion}
	\label{fig:spectogram}
\end{figure*}

\subsection{Bit-Framing}
The data packets are transmitted in small frames. Each frame consists of 46 bits and is comprised of preamble, payload, and CRC (cyclic redundancy check), as shown in Fig. \ref{fig:spectogram}.\\\\
\textbf{Preamble.} The preamble is transmitted at the beginning of every packet. It consists of a sequence of six alternating bits ('101010') which helps the receiver determine the properties of the channel, such as the carrier wave frequency and the bit period (bit rate). In addition, the preamble header allows the receiver to detect the beginning of the transmission of each packet. This is important for our covert channel, since in the case of the ultrasonic covert channel, a transmission might get interrupted, e.g., if the computer was restarted in the middle of an ongoing transmission.\\ \\
\textbf{Payload.} The payload is the 32 bits of  raw data which contains the actual packet.\\\\
\textbf{CRC.} For error detection, we insert eight bits of CRC code at the end of the frame. The receiver calculates the CRC for the received payload, and if it differs from the received CRC, an error is detected. In the case of error a packet retransmission request is sent (only in the case of bidirectional communication).

\subsection {Communication Protocol}
\begin{figure}[h]
	\centering
	\includegraphics[width=\columnwidth]{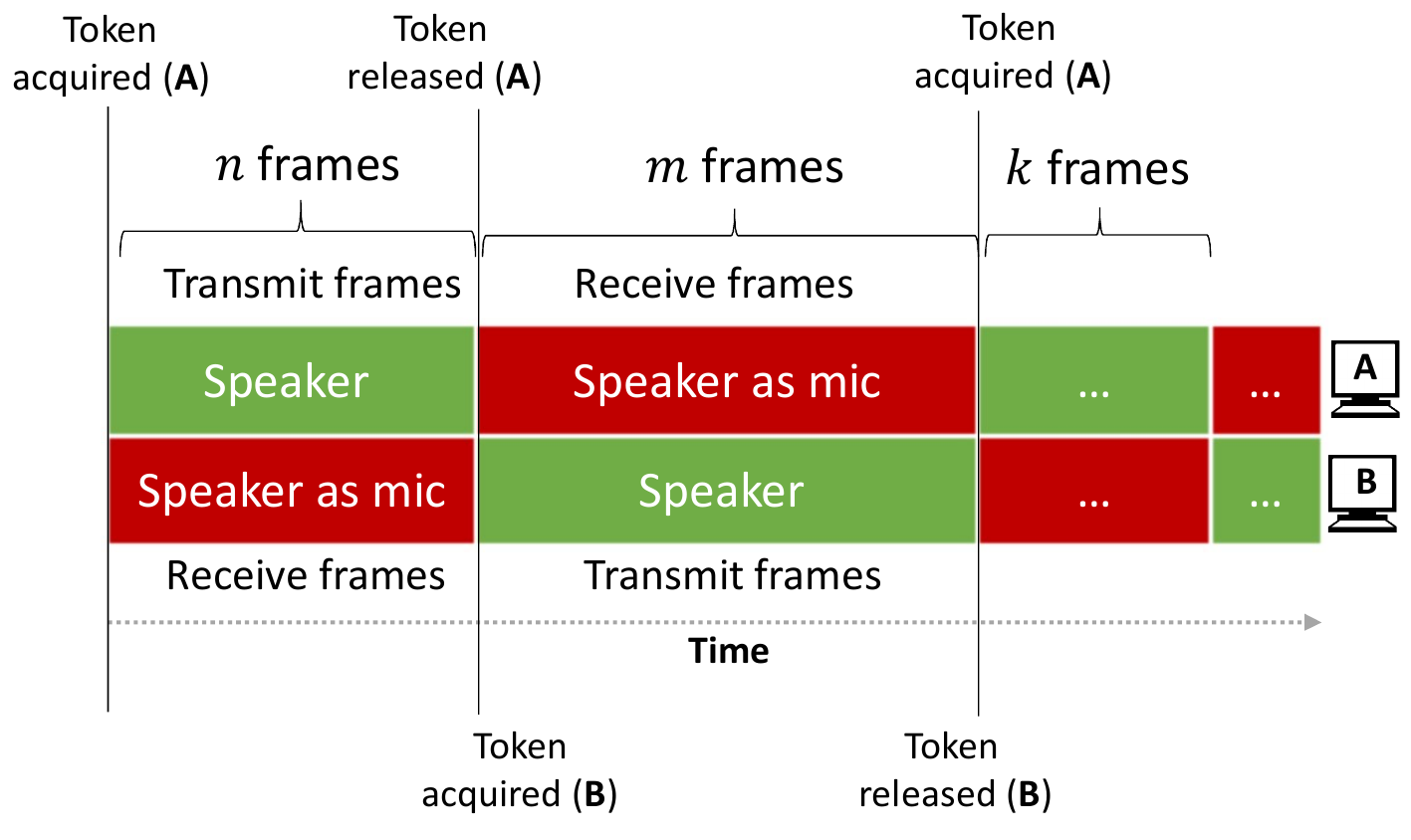}
	\caption{The communication protocol between computers \textbf{A} and \textbf{B}}
	\label{fig:protocol1}
\end{figure}

The acoustic channel is a type of shared communication channel. There are different types of multiple access protocols allowing a communication channel to be shared between many nodes (e.g., TDMA, ALOHA and CSMA \cite{rom2012multiple}). Recall that in the proposed speaker-to-speaker communication a speaker can function as either a transmitter (speaker) or receiver (microphone) at a given time. Thus, each computer must know when the speaker is being used as a speaker and when to reverse it to a microphone. We used the concept of virtual 'tokens,' in which one computer acquires a transmission token. The other computer is only allowed to transmit when a transmission token is has been released. Each computer can hold the token for a maximal time slot T\textsubscript{max}. When the computer has finished the transmission, it releases the token and begins to listen. The sequential flow of the communication between the two computers is illustrated in Fig. \ref{fig:protocol1}. At the beginning of the transmission, computer \textbf{A} acquires the transmission token, transmits $n$ frames, and releases the token. Computer \textbf{B} then acquires the token, transmits $m$ frames, and releases the token.

\subsection{Discovery Broadcast Message}
In order to establish the covert communication channel, the computers have to acknowledge each other's presence. To that end, each computer sends a broadcast message called a discovery beacon. The discovery beacon contains eight bits, which encode the computer identifier (ID). The identifier is a random number generated once at the beginning of a communication session. Since the computers are not connected, both computers might initially choose the same identifier. This case is handled by a simple rule: the first computer that detects the ID collision randomizes and broadcasts its ID. In order to discover the other computer, each computer is performs the discovery scheme outlined in Algorithm \ref{alg1}.

\begin{algorithm} 
	\caption{} 
	\label{alg1} 
	\begin{algorithmic}[1] 
		\While { (state != DISCOVERED) }
		\State $ wait(random(5000)) $
		\State $ jack\_retask(SPEAKER) $
		\State $ transmit(discoveryMessage(ID)) $
		\State $ jack\_retask(MIC) $
		\State $ message \gets waitForDiscoveryAck(5000) $
		
		\If {(message)}
		\State $ set\_state(DISCOVERED) $
		\EndIf
		\EndWhile
	\end{algorithmic}
\end{algorithm}

Each computer starts by broadcasting its ID at random times every 5000ms. Note that a random time is used in order to avoid collision with a discovery message sent by the other computer, a technique which is used in communication for collision avoidance \cite{stallings2007data}. Following the ID broadcast, the computer retasks its speaker to a microphone. It waits for a discovery acknowledgment message sent from the other computer. If an acknowledgment message is received, it stops broadcasting the discovery message.

\subsection{Type of Messages}
Table \ref{table_messages} shows the main control messages in our protocol, including the DISCOVERY, ACQUIRE and RELEASE messages described earlier in this section.  

\begin{table}[h]
	\centering
	\caption{Control messages}
	\label{table_messages}
	\begin{tabular}{lll}
		\hline
		\# & Message      & Description                                  \\ \hhline{===}
		1  & DISCOVERY    & The discovery broadcast message              \\
		2  & ACQUIRE     & Acquire the transmission token               \\
		3  & RELEASE      & Release the transmission token               \\
		4  & ACK\_OK      & Frame received successfully (ack) \\
		5  & RETRANSMIT   & Request to retransmit a frame                \\
		6  & BITRATE\_INC & Increase the current bit rate (+5\%)          \\
		7  & BITRATE\_DEC & Decrease the current bit rate (-5\%)          \\ \hline
	\end{tabular}
\end{table}

The ACK\_OK message notifies the other computer that the message was received successfully and confirms that the CRC was correct. The RETRANSMIT message requests the other computer to retransmit a frame (e.g., if the CRC is incorrect). The BITRATE\_INC and BITRATE\_DEC messages enable the two computers to agree on increasing up or decreasing down the current transmission rate in 5\%.  In particular, the  messages enable adaptive use of the channel speed to cope with environmental noise.

\subsection{Unidirectional Communication}
There are scenarios in which only one computer is equipped with a reversible speaker, while the other has an ordinary (active) speaker. In this scenario, bidirectional communication is not possible.

\begin{figure}[h]
	\centering
	\includegraphics[width=\columnwidth]{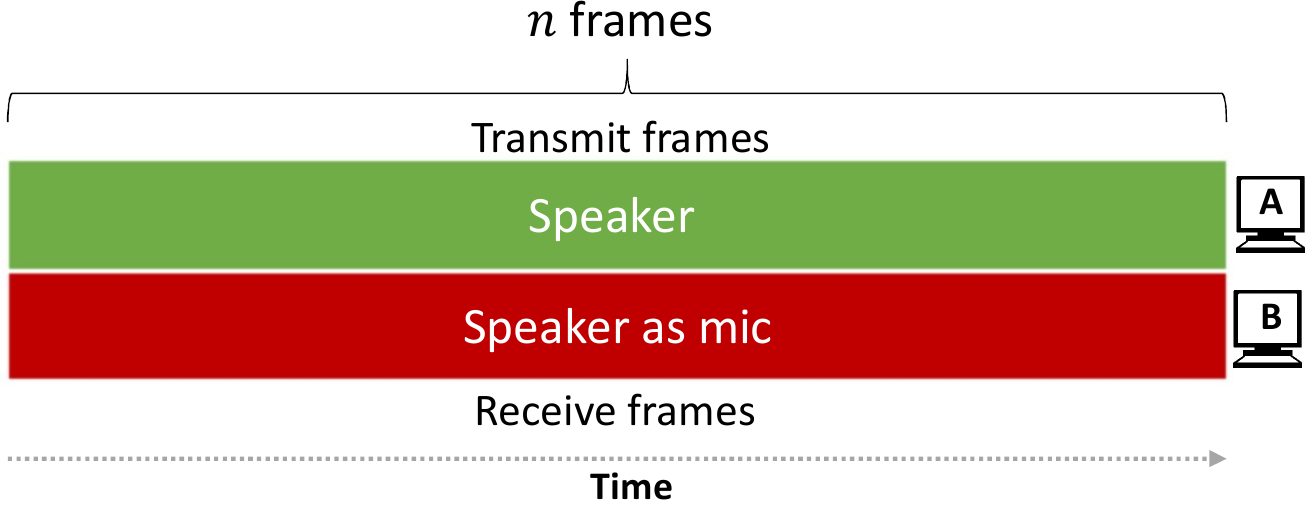}
	\caption{The unidirectional communication between computers \textbf{A} and \textbf{B}}
	\label{fig:protocol2}
\end{figure} 

This scenario is illustrated in Fig. \ref{fig:protocol2}, where computer \textbf{A} has an active (non-reversible) speaker and computer \textbf{B} has a reversible speaker. Computer \textbf{A} transmits a stream of $n$ frames to computer \textbf{B}, which receives it via the reversible speaker. Note that acknowledgments and retransmissions messages are not applicable in the case of unidirectional communication. Since in this case the transmitter and the receiver can't establish a handshake, the malware could simply be designed to initiate a data transmission and reception at a specified, predefined times (e.g, at midnight).

\section{Analysis and evaluation}
\label{sec:analysis}
Headphones, earphones, and passive speakers were not designed to perform as microphones in terms of quality and frequency range. In this section, we assess the efficacy of the speaker-to-speaker communication and present an empirical analysis of its corresponding channel capacity. We also discuss various practical considerations concerning the ultrasonic covet channel. Note that we are mainly interested in the high frequency regions that offer high channel capacity while at the same time have low auditory awareness. 

\subsection{Channel Capacity}
Channel capacity ($C$) is a measure of the theoretical upper bound on the rate at which information can be transmitted over a communication channel. We assume that $S$ is the power of the signal conveying the information and is corrupted by additive interfering Gaussian noise, with power $N$. The available communication bandwidth is $B$ (in Hz). Given that, the channel capacity in bits per second can be calculated using the Shannon-Hartley theorem: \\
\begin{equation} \label{eu_eqn}
C=Blog_2(1+\frac{S}{N})
\end{equation}
\\
Intuitively, this formula informs us that the higher the signal-to-noise ratio (SNR), and channel bandwidth, the higher the amount of information that can be conveyed.\\ 

We calculate the capacity of a communication channel formed between two loudspeakers, one of which serves as a transmitter and the other serves as a receiver. In these experiments, a sweep sinusoid of ten second in length at a range of 1Hz to 24kHz is played from the transmitter and recorded by the receiver. We use the Praat \cite{Praatdoi67:online} tool to perform a short-time spectral analysis of the received signal.\\\\
\textbf{Measurement setup.} We evaluate the channel capacity for distances of one, four and eight meters between the transmitter and the receiver. To that end, we tested three off-the-shelf passive loudspeakers as receivers:  (1) Logitech Z523, (2) Logitech Z213, and (3) Philips SPA5300. We also tested a pair of small Samsung earbuds for comparison. The loudspeakers were connected to a retaskable audio output jack on an Optiplex 9020 desktop PC. The sweep signal was played through a Logitech Z100 loudspeaker connected to a Gigabyte GA-H97M-D3H desktop workstation, (Intel Core i7-4790) running Ubuntu 16.04.1 kernel 4.4.0.\\\\
\textbf{Calculations.} The signal is analyzed in successive Gaussian windows of 200 milliseconds with 25\% overlap in time.  We adopt a frequency resolution of 100Hz for each band, resulting in 250 analyzed bands. The SNR is estimated for each frequency band, as the power ratio of the received signal and the measured noise in this band. \\

\begin{figure}[t]
	\centering
	\includegraphics[width=\columnwidth]{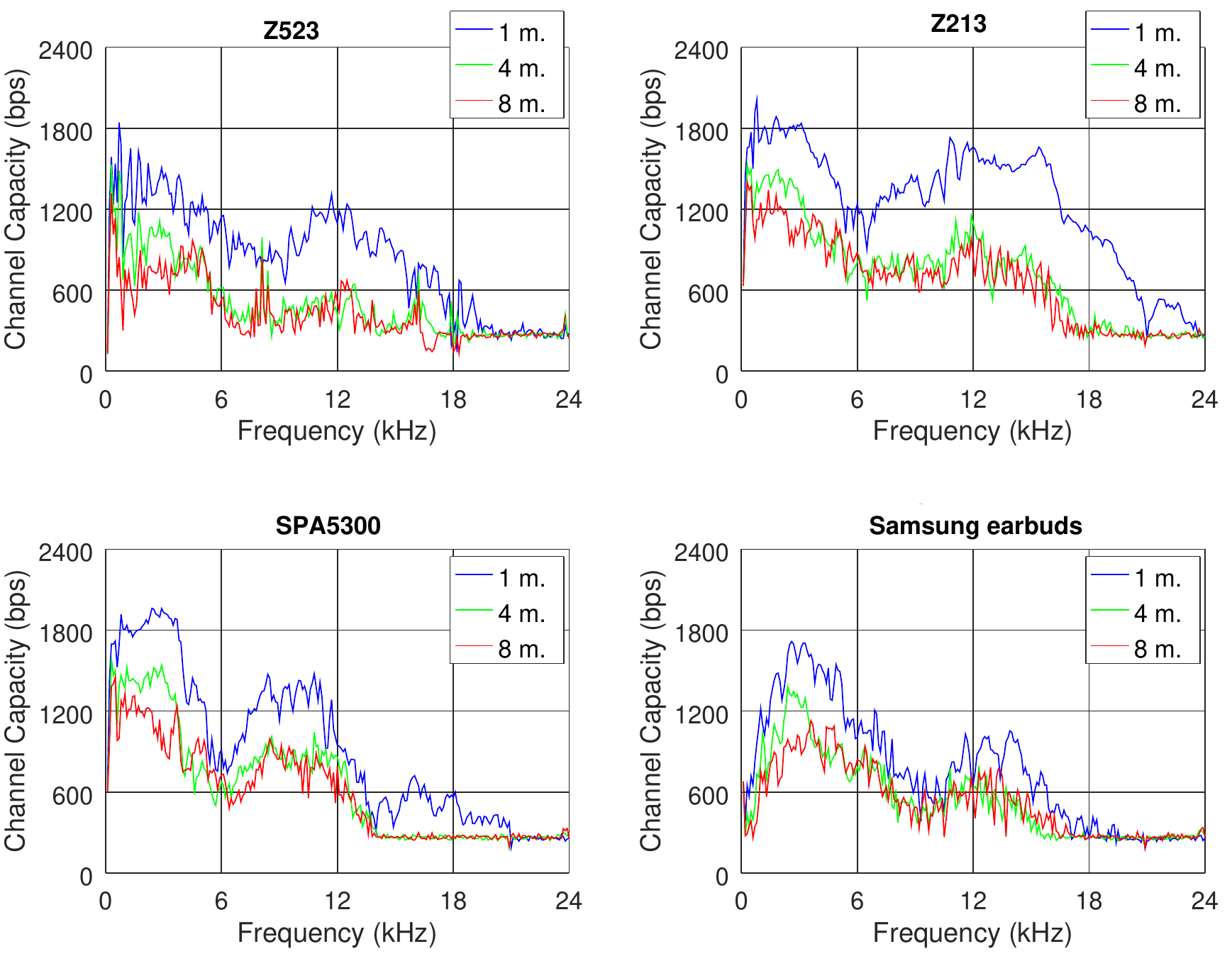}
	\caption{Channel capacity of speaker-to-speaker communication}
	\label{fig:s2s}
\end{figure} 

Fig. \ref{fig:s2s} presents the evaluated channel capacity for the entire frequency range.  We can observe that for the 1m, 4m and 8m setups, the theoretical upper bound for the channel capacity is between ~1200 bit/sec and ~1800 bit/sec for the audible frequency bands (lower than 18kHz). As expected, the channel capacity is correlated with the distance between the transmitter and the receiver.  The channel capacity significantly degrades in the sub-bass range (up to about 60Hz) and for high frequencies (above 18kHz). In these ranges, the theoretical upper bound is between ~300 bit/sec and ~600 bit/sec in most cases. The reason for that is that loudspeakers, and particularly home grade PC loudspeakers, were projected and optimized for human auditory characteristics, and therefore they are more responsive to the audible frequency ranges. \\

\subsubsection{Headphones, Earphones and Earbuds}
We also calculate the capacity of a communication channel in which headphones, earphones and earbuds are used as receivers. Similar to the previous experiment, we sweep sinusoid of ten-second length in a range of 1Hz to 24kHz and record it by the headphones. We evaluate the channel capacity for one, five and eight meters. We tested four types of headphones (1) Philips vibes earbuds, (2) Philips SHS3300 earhooks, (3) Logitech h110 headphones and (4) Philips SHL3850NC headphones. The headphones were connected to a retaskable audio output jack on the desktop PC described above.

\begin{figure}[t]
	\centering
	\includegraphics[width=\columnwidth]{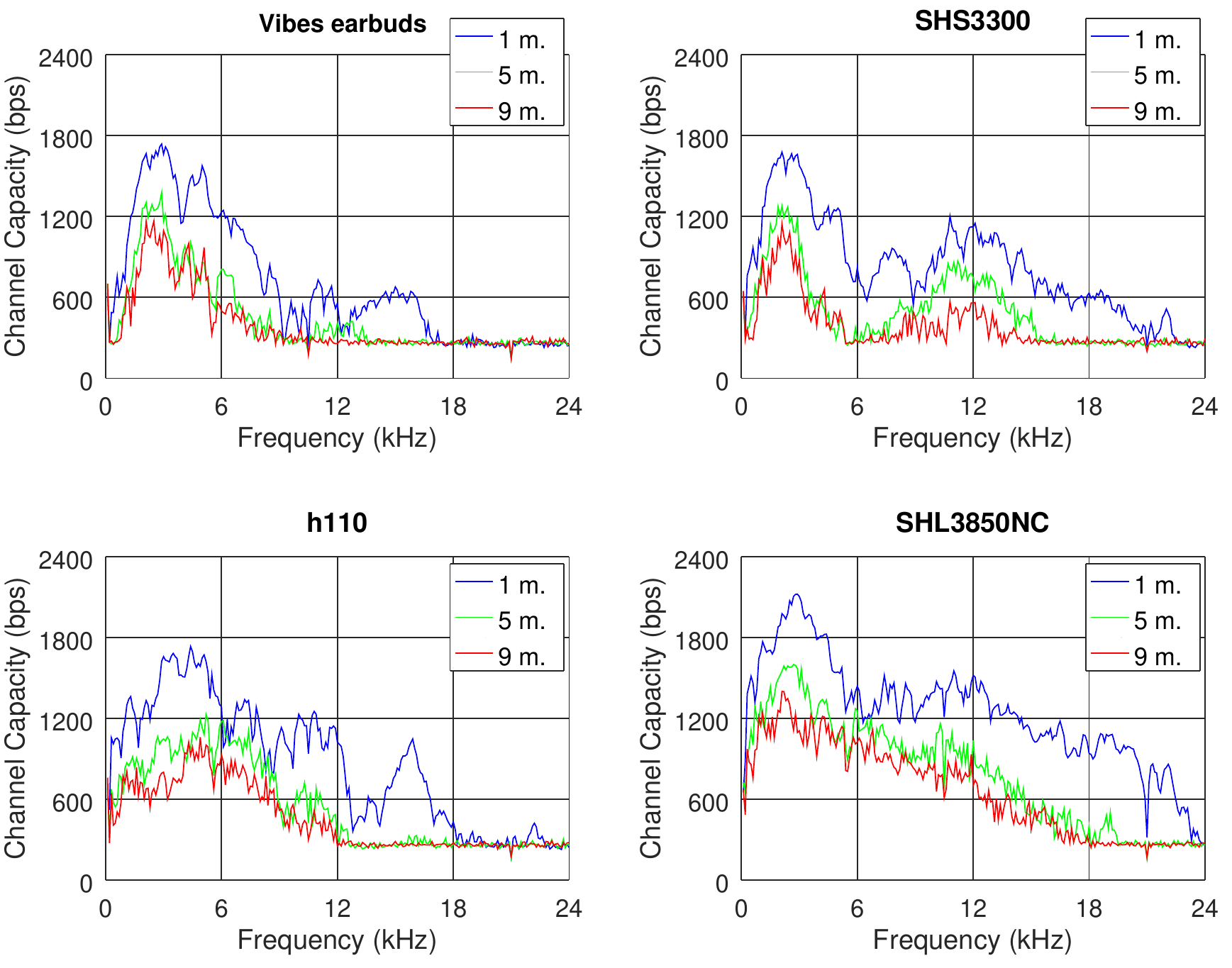}
	\caption{Channel capacity of speaker-to-headphones communication}
	\label{fig:s2h}
\end{figure} 

The results are shown in Fig. \ref{fig:s2h}. We observed that loudspeakers do not perform significantly better as receivers than earbuds or headphones, as one could expect. In particular, for the 1m, 5m, and 8m setups, the theoretical upper bound for the channel capacity is between 300 bit/sec and 600 bit/sec in most cases. \\
\subsubsection{Headphones-to-Headphones Communication}

\begin{figure}[t]
	\centering
	\includegraphics[width=0.7\columnwidth]{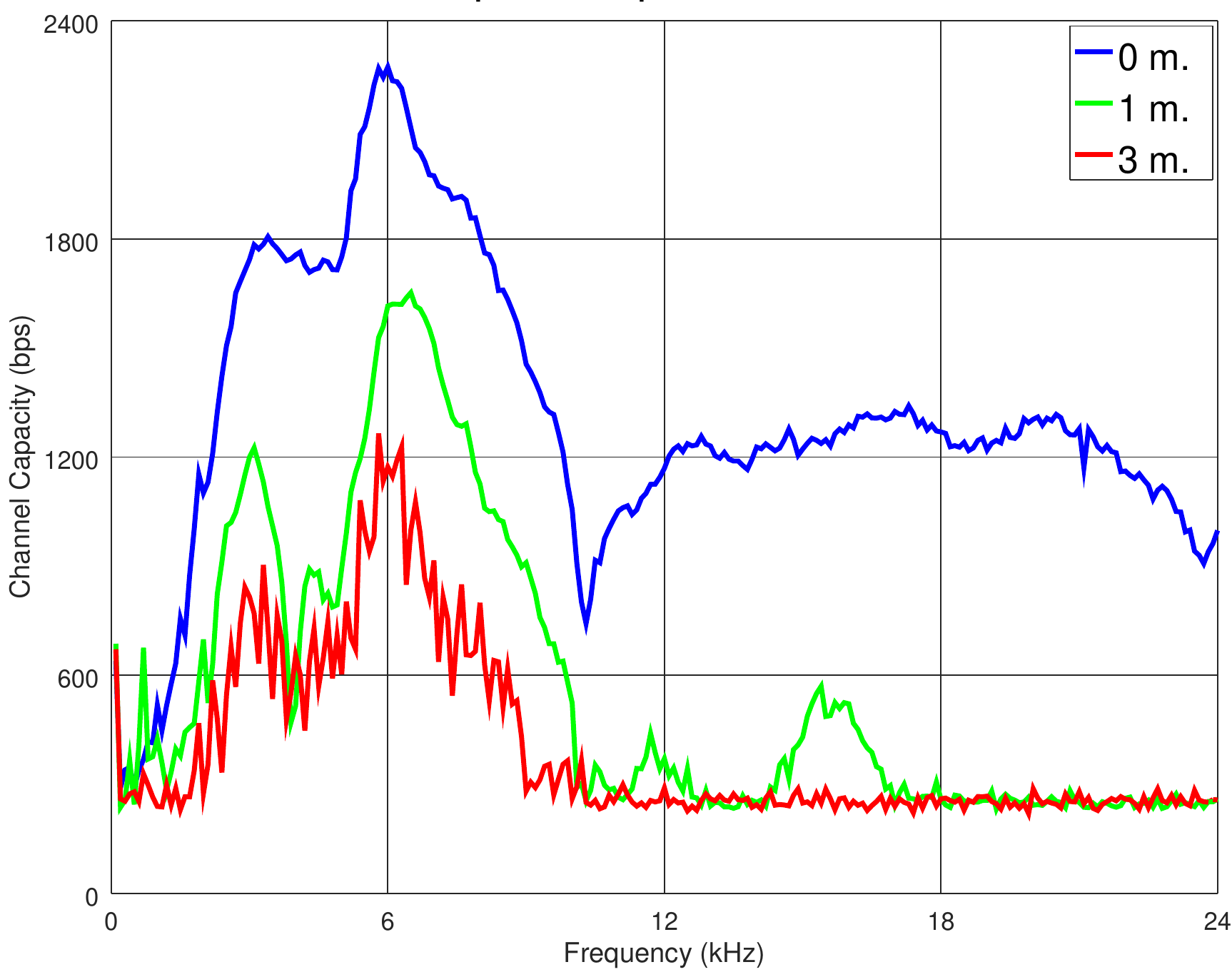}
	\caption{Channel capacity of headphones-to-headphones communication}
	\label{fig:h2h}
\end{figure} 

To complete the whole picture, we demonstrate the effects of using headphones as both the transmitter and the receiver. The test signal was played through the h110 headphones and captured by the SHL3850NC headphones. Fig. \ref{fig:h2h} presents the evaluated channel capacity for the entire frequency range.  The results indicates that the headphone-to-headphone communication is limited to about three meters. The channel capacity at high frequencies (above 18kHz) is limited to ~250 bit/sec. In the context of the attack model, this implies that headphones-to-headphones communication is relevant only in certain cases, e.g., where the headphones are located side by side, or on two adjacent tables.
\subsection{Practical Considerations}
In this sub-section, we discuss the practical considerations concerning the ultrasonic covert channel. We examine the effect of environmental noise on the channel, the equipment's position and the feasible transmission rates in a typical working place.
\subsubsection{Environmental Noise (music and speech)}

\begin{figure}[t]
	\centering
	\includegraphics[width=0.75\columnwidth]{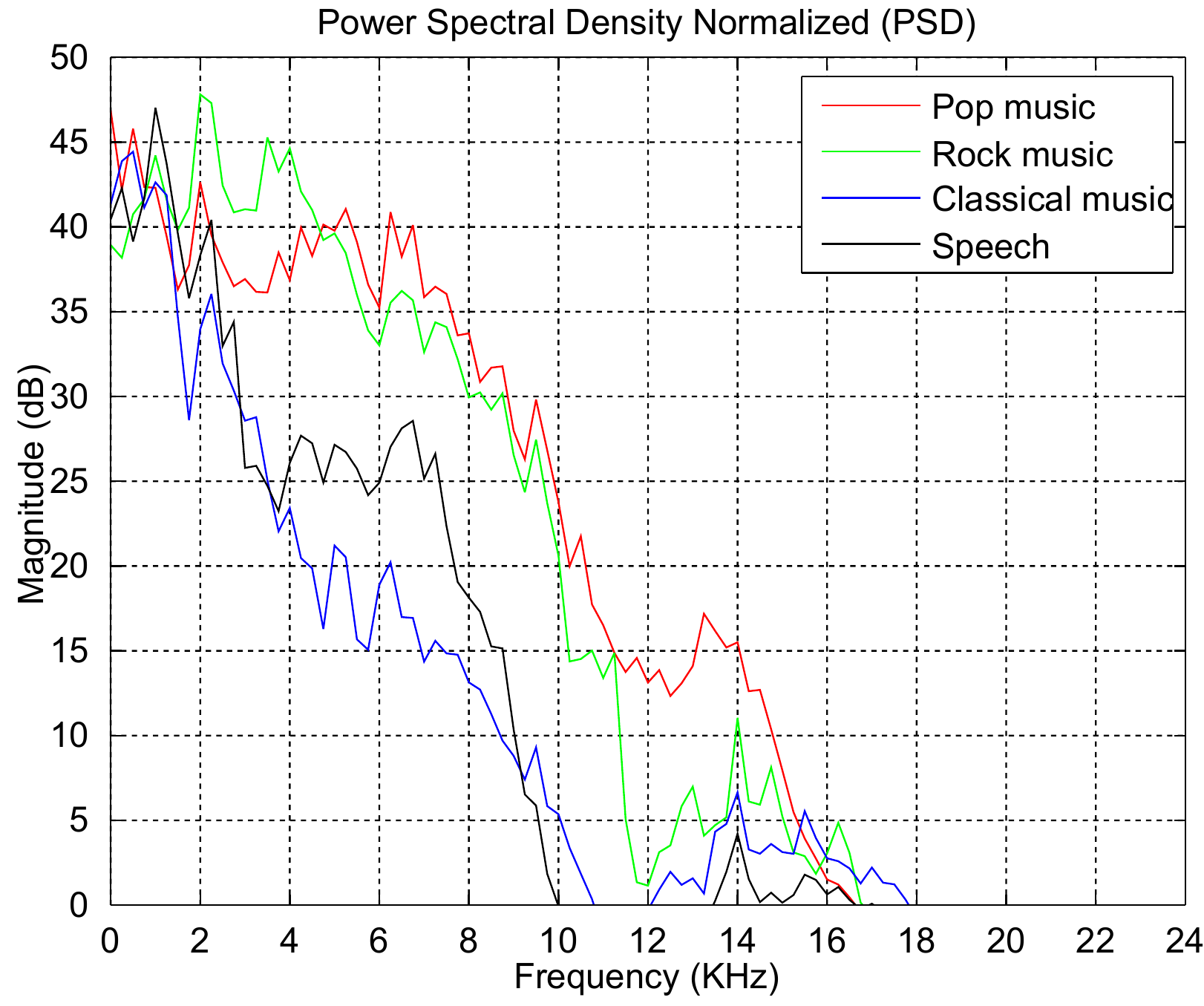}
	\caption{The power spectral density (PSD) of various types of music and speech}
	\label{fig:PSD}
\end{figure} 

We start by examining a situation in which the covert channel is employed in a setting in which there is an interfering noise signal. For instance, when music is being played or people are talking in the room. In this case, our channel capacity might be decreased due to the SNR conditions. We demonstrates the background noise scenario by playing a series of high definition (HD) music clips in the room. The series includes pop, rock and classical music clips randomly chosen from YouTube. For human speech we played Bill Gates' speech delivered at Harvard University\footnote{Bill Gates' speech at Harvard University (https://www.youtube.com/watch?v=3bDqJp-NgF4)}. 
Fig. \ref{fig:PSD} shows the normalized power spectral density (PSD) of the interfering music. The PSD shows how the power of the generated signals is distributed over the entire frequency band (1Hz-24kHz). It can be seen that although the interfering noise spreads throughout the whole frequency band, a very small amount of energy is concentrated above 18kHz. The same is true for interfering speech, since it is narrow banded in comparison to music. Moreover, the human speech intensity is highly concentrated at relatively low frequency bands. The speech of an adult male has a fundamental frequency ranging from 85Hz to 180Hz, and that of an adult female ranges from 165Hz to 255Hz. The spectral view shows that a covert transmission above 18kHz would experience less interference from background music or human speech in the room.

\begin{figure*}[t]
	\centering
	\includegraphics[width=0.67\textwidth]{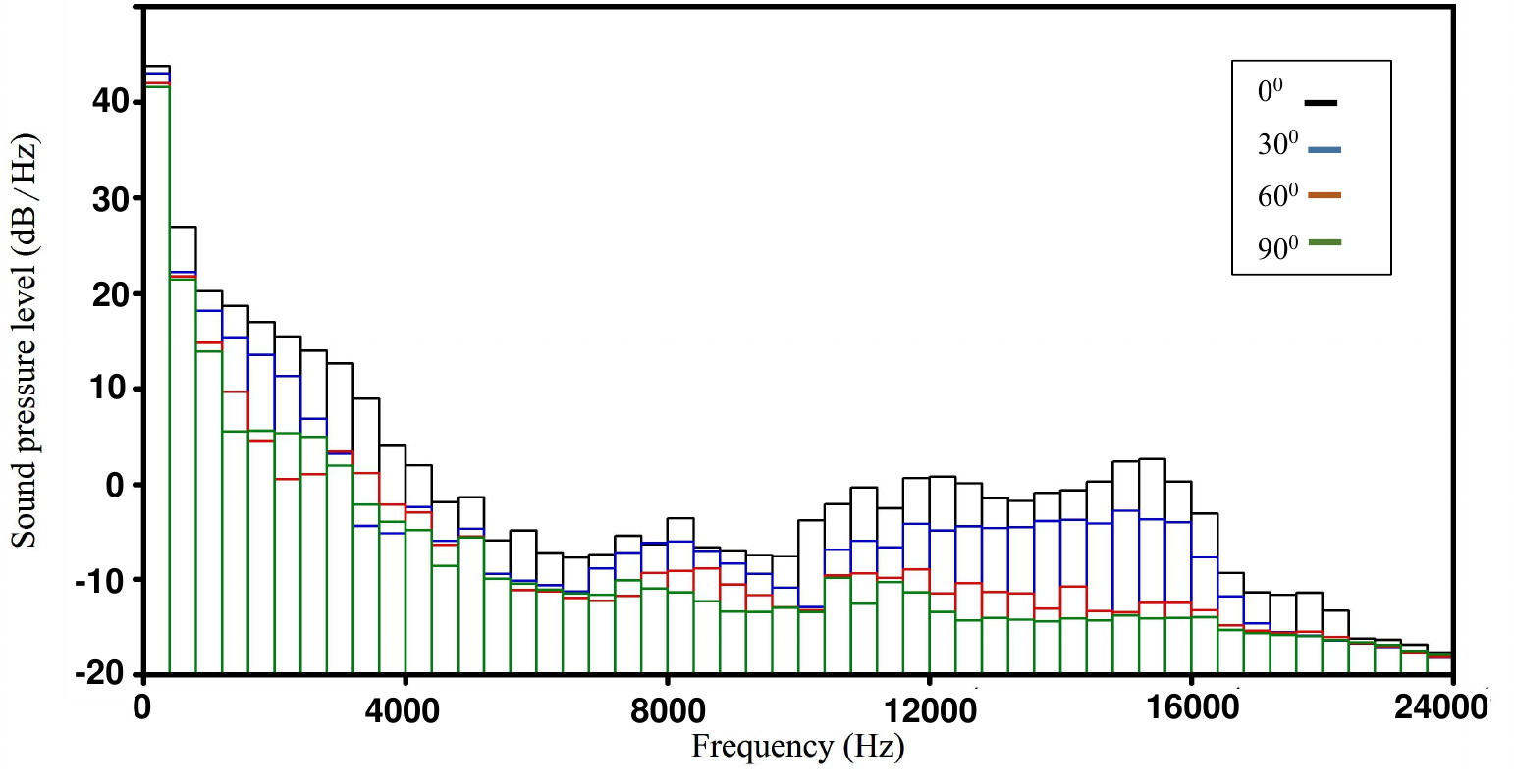}
	\caption{The frequency response in different dispositions of the transmitter and the recevier}
	\label{fig:angle}
\end{figure*}

\subsubsection{Positions}
The communicating transmitters and receivers might be positioned in various layouts and directions. In the acoustic channel, the position of the transmitters and receivers significantly affects the quality of the received signal \cite{fahy2000foundations}. Specifically, the SNR degrades when the transmitter and receiver speakers are not aligned. In acoustics, directivity describes the way a speaker's 
(or microphone's) frequency response changes at off axis angles \cite{tucker1966applied}. A \textit{wide directivity} speaker maintains the signal quality consistency between the on and off axis while \textit{narrow directivity} speaker is one where the signals quality is substantially different between the on and off axis. The computer loudspeakers are of narrow directivity, and hence, they loose off axis response at lower frequencies compared to the on axis response. This phenomenon is called "beaming" and intuitively corresponds to the sensation of frequency unbalance experienced when one moves from side to side across a speaker \cite{mccowan2001microphone}. Beaming affects higher frequencies more than the lower spectrum. In theory, off-axis begins to affect the response at frequencies having a wavelength close to the diameter of the radiating cone. The approximate starting beaming frequency $f$ is provided by:

\begin{equation} \label{eu_eqn}
f \approx c/D 
\end{equation}

where $c$ is the speed of sound (~340 m/s) and $D$ is the diameter of the speaker cone. Therefore, considering a PC speaker having a 10cm. cone diameter, beaming will start at approximately  3400Hz. In practice, the geometry of the radiation cone and other factors cause beaming to start at  lower frequencies, as was observed in our experiments. Fig. \ref{fig:angle} displays the spectrum of a sweep signal received by a reversed loudspeaker from different angles with regard to the transmitter speaker. As expected, the off-axis response at 30, 60 and 90 degrees significantly decreases for increasing angles. The SNR degradation is visibly stronger at high frequencies. Interestingly, due to their reduced cone diameter, headphones and earbuds in transmitting mode are less affected by beaming.

\subsubsection{Bit Error Rate}
The transmission rates of the ultrasonic covert channel have been extensively measured in several prior work \cite{hanspach2014covert,carrara2014acoustic,deshotels2014inaudible}. In this research, we aim at examining the practical considerations of the covert channel and the corresponding transmission rate with the speaker-to-speaker communication. That is, we measure the transmission rates that yield low bit error rates (\~1\%)  during the transmissions. Note that the channel capacity discussed earlier represents the upper theoretical limits of the communication channel. The actual bit rate is usually lower than the channel capacity and is determined by the modulation scheme and the quality of the transmitter and receiver used. Our experiments shows that at a distance of three meters between two speakers (Z523 and Z213), a transmission rate of 166 bit/sec results in a 1\% bit error rate, during the exfiltration of a 1Kbit binary file. However, at distances of 4-9 meters, the 1\% bit error rate is only achieved at transmission rates of 10 bit/sec. Our waveform analysis shows that the signal quality is degraded at distances greater than four meters mainly due to the environmental noise, which results in a lower SNR.

\begin{table*}[t]
	\centering
	\caption{Defensive Countermeasures}
	\label{table_cm}
	\begin{tabular}{lll}
		\hline
		Countermeasure                                                                                 & Advantages                                                          & Limitations                                                                                              \\ \hline
		Prohibit the use of headphones/earphones/speakers                                              & \begin{tabular}[c]{@{}l@{}}Hermetic protection\end{tabular} & Poor usability                                                                                     \\
		Use active speakers / on-board amplifiers                                                    & \begin{tabular}[c]{@{}l@{}}Hermetic protection\end{tabular} & Not relevant for headphones and earphones                                                          \\
		\begin{tabular}[c]{@{}l@{}}Disable audio codec in BIOS/UEFI \end{tabular}                              & \begin{tabular}[c]{@{}l@{}}Easy to deploy\end{tabular}      & \begin{tabular}[c]{@{}l@{}}Poor usability\end{tabular}                                           \\
		
		\begin{tabular}[c]{@{}l@{}}Detect jack retasking / enforce jack retasking policies \end{tabular}                            & \begin{tabular}[c]{@{}l@{}}Easy to deploy\end{tabular}      & \begin{tabular}[c]{@{}l@{}}Can be evaded by advanced malware \& rootkits\end{tabular}                          \\
		\begin{tabular}[c]{@{}l@{}}Use ultrasonic noise emitters (signal jamming)\end{tabular} & \begin{tabular}[c]{@{}l@{}}Generic solution\end{tabular}    & \begin{tabular}[c]{@{}l@{}}Hard to deploy due to the noise generated\end{tabular} \\
		\begin{tabular}[c]{@{}l@{}}Detect ultrasonic transmission (signal detection)\end{tabular} & \begin{tabular}[c]{@{}l@{}}External (non-invasive) \end{tabular}    & \begin{tabular}[c]{@{}l@{}}Reliability \end{tabular} \\
		\begin{tabular}[c]{@{}l@{}}Low-pass filters (software/hardware)\end{tabular} & \begin{tabular}[c]{@{}l@{}}Generic solution\end{tabular}    & \begin{tabular}[c]{@{}l@{}}Deployment and additional cost (hardware filters)\end{tabular} \\
		\hline
	\end{tabular}
\end{table*}

\section{Countermeasures}
\label{sec:counter}
Countermeasures can be categorized into hardware and software countermeasures.

In highly secure facilities it is common practice to forbid the use of any types of loudspeakers (passive or active) to create so-called audio-gap separation between computers \cite{Jumpingt83:online}. Less restrictive policies prohibit the use of microphones but allow one-way loudspeakers. Such a policy was suggested by the NSTISSAM TEMPEST/2-95, RED/BLACK guide \cite{NSTISSAM75:online}. In this guide the protective measures state that \textit{"Amplifiers should be considered for speakers in higher classified areas to provide reverse isolation to prevent audio from being heard in lesser classified areas."} Accordingly, some TEMPEST certified loudspeakers are shipped with amplifiers and one-way fiber input \cite{TEMPESTV49:online}. However, the aforementioned policies and protective countermeasures are not relevant to most modern headphones, which are primarily non-powered, and built without amplifiers. A general solution for all kinds of speakers and headphones is to implement the amplifier on-board, integrating it within the audio chipset. 

A different approach is to mask ultrasonic transmissions in certain area by using ultrasonic jammers. These devices generate ultrasonic background noise aimed at interfering with the covert communication signals.  \cite{9Counter84:online}. Note that this type of solution is not trivial to deploy on a wide scale since the  jamming range is limited to a radius of a few meters to a single room. The jamming efficacy also depends on the distance from the potential transmitters and receivers. Carrara \cite{carrara2014acoustic} suggested monitoring the audio channel for abnormally peaks of energy, in order to detect hidden transmissions in the area. In our case, the ultrasonic frequency range above 18kHz should be scanned (continuously) and analyzed. However, as noted in \cite{carrara2014acoustic}, if the hardware device scanning the ultrasonic spectrum is far from the transmitter this approach may not be effective.

Software countermeasures include completely disabling the audio hardware in the UEFI/BIOS. This can prevent a malware from accessing the audio codec from the operating system level. However, such a configuration eliminates the use of the audio hardware (e.g., for playing audio), and hence, may not be feasible in all cases. Another option is to install a HD audio driver that prevents jack retasking or enforces a strict jack retasking policy. To provide general software-level protection, anti-malware and intrusion detection systems can employ a monitoring driver which detects unauthorized speaker-to-mic retasking operations and block them. Another approach proposed by \cite{hanspach2014covert} is to filter out the inaudible frequencies at the range of 18kHz and higher with a low-pass or bandpass filter. Recently, a software based ultrasonic firewall (dubbed SilverDog) was implemented for the Google Chrome browser \cite{GitHubub23:online}. This open-source project aims at blocking cross-device tracking which utilizes ultrasonic beacons (uBeacons) \cite{mavroudis2017privacy}. 

\begin{figure}[h]
	\centering
	\includegraphics[width=0.7\columnwidth]{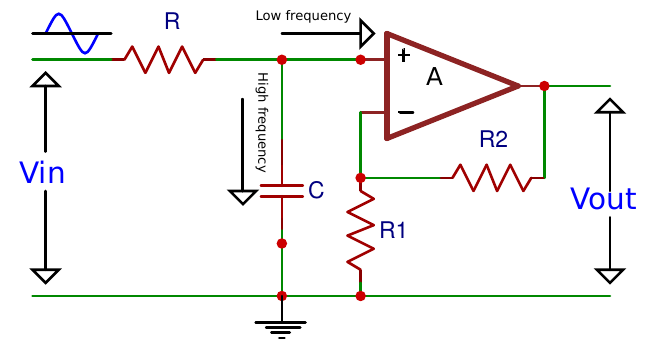}
	\caption{Low-pass filter circuit for 3.5mm audio jack}
	\label{fig:lowpass}
\end{figure}

To prevent a malware initiated ultrasonic covert channel, the filter could be implemented as an audio filter (or 'mixer') in the operating system. The main drawback of this approach is that it can be disabled or bypassed by advanced malware and rootkits. 
For an increased level of protection, we implemented the low-pass filter as a \textit{trusted} component in hardware. Fig. \ref{fig:lowpass} shows the circuit design of a low-pass filter with an amplifier, for a 3.5mm audio jack. Note that the cutoff frequency of in this filter is determined by the capacitor \textit{C} and the resistor \textit{R}.
In our case, the circuit pass signals with a frequency lower than 18kHz and attenuates signals with frequencies higher than 18kHz. For technical information on low-pass filters and their functionality, we refer the interested reader to relevant textbooks in this topic \cite{sedra1998microelectronic}.   
The countermeasures are listed and summarized in Table \ref{table_cm}.

\section{Conclusion}
\label{sec:conclusion}
It is known that covert communication can be established between two nearby air-gapped computers, enabling them to communicate to one another via ultrasonic waves \cite{hanspach2014covert}. However, the standard attack model requires the two computers to be equipped with both speakers and microphones. Consequently, this type of covert channel is not applicable in secure facilities where it is common practice to prohibit the use of microphones \cite{Jumpingt83:online}. Also, many desktop workstations lack microphones or the microphones have been physically muted or turned off \cite{WhyhasMa55:online}.  
In this work, we show how air-gapped computers without microphones can still exchange data via ultrasonic waves. The computers must be connected to passive speakers, headphones, or earphones. Our method is based on the capability of a malware to transform a PC's connected speaker from an output device to an input device, unobtrusively changing its role from a speaker to a microphone \cite{guri17speake}.  We show that although the reversed speakers are not designed to function as microphones, they are still sensitive to high frequency sound waves at a range of 18kHz to 24kHz.
Transmissions in this range are practically inaudible to most adults, and hence this channel is considered covert. We evaluate the communication channel and present three attack scenarios: (1) speaker-to-speaker communication, (2) speaker-to-headphones communication, and (3) headphones-to-headphones communication. Our results show that by using loudspeakers, data can be exchanged over an air-gap from a distance of eight meters away with an effective bit rate of 10 - 166 bit/sec. When using two headphones, the distance is limited to three meters away. This enables 'headphones-to-headphones' covert communication, which is discussed for the first time.

\bibliographystyle{plain}
\bibliography{sts}

\end{document}